\documentclass{emulateapj}

\newcommand{\ebv}{$E(B-V)$}

\newcommand{\z}{$z$}

\newcommand{\zs}{$z$$\sim$}

\newcommand{\hdffilters}{$U_{300}B_{450}V_{606}I_{814}J_{110}H_{160}$}

\newcommand{\chisq}{$\chi^2$}

\newcommand{\SEDfit}{\texttt{SEDfit}}
\newcommand{\makesed}{\texttt{make\_sed}}
\newcommand{\fitsed}{\texttt{fit\_sed}}

\shorttitle{SEDfit}
\shortauthors{Sawicki}

\begin{document}


\title{SEDfit: Software for Spectral Energy Distribution Fitting of Photometric Data}

\author{Marcin Sawicki}
\affil{ Department of Astronomy and Physics and Institute for Computational Astrophysics, Saint Mary's University, 923 Robie Street, Halifax,
Nova Scotia, B3H 3C3, Canada}
\email{sawicki@ap.smu.ca}

\slugcomment{PASP, in press}

\begin{abstract}
This paper describes \SEDfit, the earliest --- but continually upgraded --- software package for spectral energy distribution fitting (SED fitting) of high-redshift photometric data, and the only one to properly treat non-detections. The principles of maximum-likelihood SED fitting are described, including formulae used for fitting both detected and un-detected (upper limits) photometric data. The internal mechanics of the \SEDfit\ package are presented and several illustrative examples of its use are given. The paper concludes with a discussion of several issues and caveats applicable to SED-fitting in general. 
\end{abstract}


\section{INTRODUCTION}\label{sec.introduction}

The study of how galaxies form and evolve is one of the most active fields of astrophysical research today. Key in this work is the ability to estimate the stellar masses, star formation rates (SFRs), and dust content of faint galaxies in the distant Universe. 

One of the most widely-used approaches to determining these quantities has been through the quantitative comparison of observed multi-wavelength spectral energy distributions (SEDs) with theoretical or empirical models.  This approach was first developed for high-$z$ galaxies by Sawicki \& Yee (1998) and has subsequently been adopted and elaborated on by numerous other studies (e.g., Ouchi, Yamada, Kawai, \& Ohta 1999; Papovich et al. 2001; Hall et al.\ 2001; Shapley et al.\ 2001; and, subsequently, many others).  It is fair to say that SED fitting has now become a standard way to estimate stellar masses and other properties of high-redshift galaxies.  

A related task is the estimation of redshift from broadband photometry, a technique known as photometric redshifts (photo-$z$'s). When photo-$z$ estimation is done by means of the template fitting approach (Gwyn \& Hartwick 1996; Sawicki, Lin, \& Yee 1997), it can be regarded as a subset of the information that can be routinely extracted from full SED-fitting.  Indeed, photo-$z$ estimation by template fitting has led to the subsequent development of SED-fitting as a tool for high-redshift galaxy studies. 

This paper gives a description of the \SEDfit\ software tool, which is the first  --- though continually-evolving --- among many software packages now used in SED-fitting work; it is also the only such package at present to properly treat upper limits --- i.e., observed but undetected photometric data.  To date, \SEDfit\ has been used to estimate or simulate photometric redshifts (Sawicki, Lin, \& Yee, 1997; Hogg et al.\ 1998; Sawicki 2002; Sawicki \& Mall\'en-Ornelas 2003; Sorba \& Sawicki 2010; Sorba \& Sawicki 2011), to determine the physical properties of distant galaxies such as their stellar mass, starburst age, and reddening (Sawicki \& Yee 1998; Sawicki 2001; Yabe et al.\ 2009; Yuma et al. 2010, 2011; Hashimoto et al.\ 2010; Palmer 2010, 2012), or to do both at the same time (Thompson et al.\ 1999; Hall et al.,\ 2001; Sawicki 2012). It has also been used for simpler, but nevertheless important, tasks such as calculating colors and k-corrections of distant galaxies (Sawicki et al.\ 2005; Sawicki \& Thompson 2006a, 2006b, Arcila-Osejo, Sawicki, \& Sato 2012).  It continues to be used in these and similar roles. 

The paper is organized as follows.  Section~\ref{sec.overview} presents an overview of \SEDfit, while subsequent sections delve into some of the key details: \S\ref{sec.flux-generation} describes the generation of grids of model SEDs and \S\ref{sec.fitting} describes the fitting procedure by means of which the observational data are compared to these SED grids to identify the best-fitting model. The detailed derivation of the fitting formalism, including the formulae for the case when some of the photometric observations yield only non-detections, is given in the Appendix. Section~\ref{sec.examples} gives several examples of the application of the \SEDfit\ software and \S\ref{sec.discussion} discusses several issues associated with SED fitting in general.

\section{OVERVIEW AND IMPLEMENTATION}\label{sec.overview}

\begin{figure*}
\begin{center}
\includegraphics[width=0.85\textwidth]{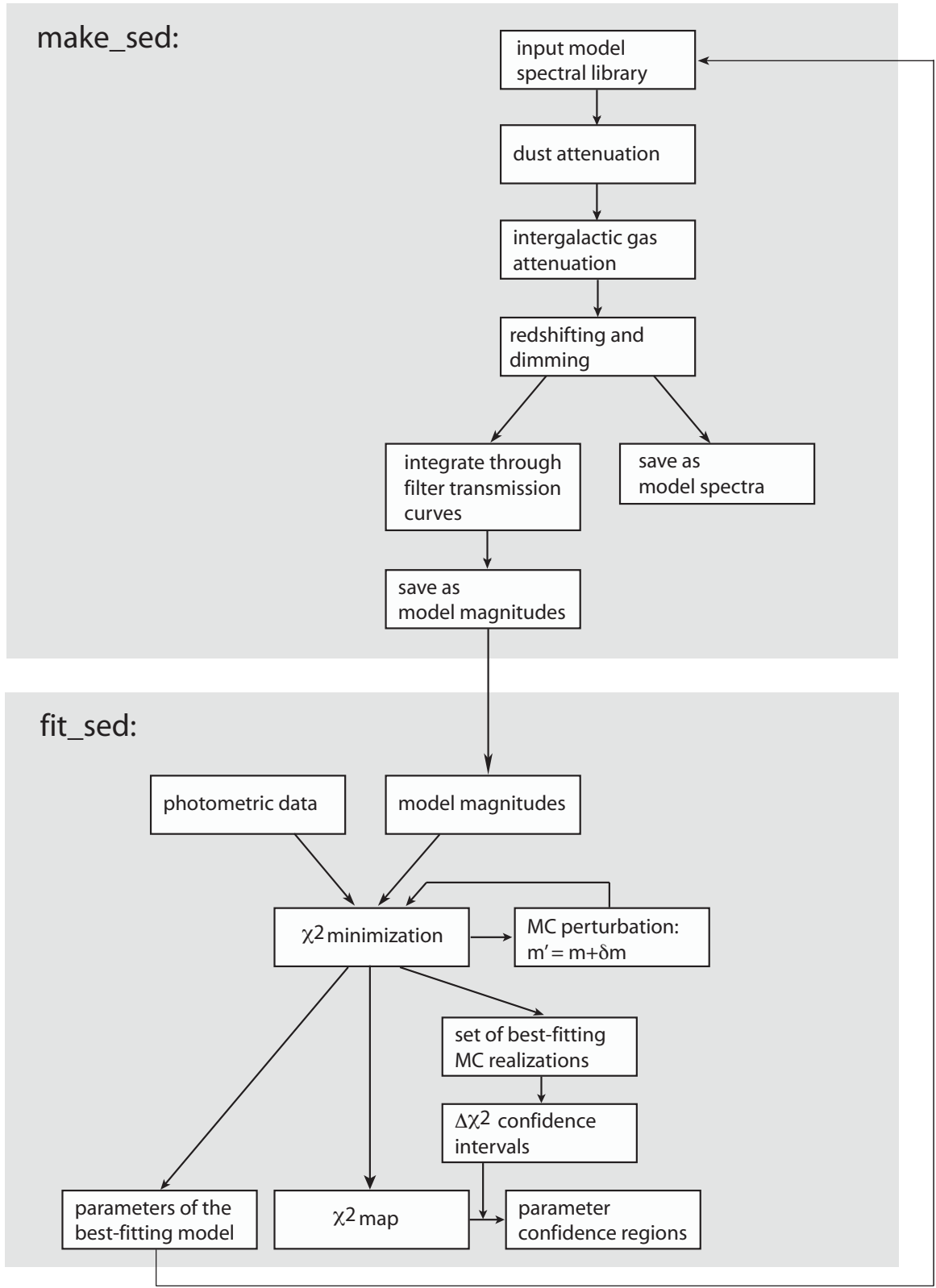} 
\caption{
\label{fig.flowchart} 
Flowchart of \SEDfit\ operations.  The upper part of the diagram illustrates the operations performed by \makesed, the program that generates model spectral energy distributions.  The lower part shows the operations performed by \fitsed, the program that performs the comparisons between models and data.  
}
\end{center}
\end{figure*}

Given a distant galaxy, the task at hand is to compare a set of photometric observations made through (usually) broadband filters with sets of flux densities predicted for a grid of spectral models, in order to identify that model which best matches the data and to establish the degree of confidence in that model. The task naturally separates into two distinct components: (1) the generation of model flux densities, and (2) the quantitative comparison of these model flux densities to the observational data.  \SEDfit\ deals with these two components by means of two separate software programs: \makesed\ and \fitsed\ (Fig.~\ref{fig.flowchart}).

The program \makesed\ carries out the tasks associated with the first of these two major components, namely the generation of a grid of model magnitudes (or spectra). It starts by taking as input a set of rest-frame model spectra (from, e.g., the GALEXEV library of Bruzual \& Charlot 2003) and performs the following operations: attenuation by interstellar dust, cosmological dimming and redshifting to the observed frame, and redshift-dependent attenuation by intervening intergalactic gas clouds (see top half of Fig.~\ref{fig.flowchart}). The spectra thus adjusted can then be saved, or can be further integrated through desired system transmission curves to produce grids of model magnitudes. The details of this process are described in \S\ref{sec.flux-generation}. The program \makesed\ loops over the full specified ranges for all the user-specified parameters (redshift, reddenning, etc.), as well as parameters internal to the input spectral templates (such as stellar population ages) to produce grids of models as a function of these parameters. 

Next, the program \fitsed\ takes the grid of model magnitudes generated by \makesed\ and compares it to the observed photometry to find the model that best matches the data.  The comparison is done in flux space (after converting from magnitudes to flux densities) by means of a maximum likelihood test (``$\chi^2$ fitting") modified to correctly account for those photometric observations that have only yielded upper limits. The details of the mathematical formalism that underlies the fitting procedure used by \fitsed\ are given in \S\ref{sec.fitting}.  In addition to identifying the best-fitting model and providing the corresponding parameters and their uncertainties, \fitsed\ can optionally also produce \chisq\ maps of the parameter space, Monte Carlo simulations of the uncertainties, and/or a means to produce the model spectra that correspond to the best-fitting model. The program can loop over multiple objects in a catalog, fitting each one with the full grid of models or a subset thereof. 

The present implementation of the software is in Perl and the Perl Data Language (Glazebrook \& Economou 1997). While Perl itself is a scripting rather than a compiled language, the Perl Data Language module is very efficient at dealing with large arrays because it uses pre-compiled libraries. The computationally-intensive work in \SEDfit\ is done using this Perl Data Language functionality and is therefore quite fast.  In particular, array operations are very efficient in Perl Data Language, making it possible for SEDfit to perform much of the SED generation and fitting very rapidly. 

Both \makesed\ and \fitsed\ are operated from the Unix command line and are controlled by means of editable parameter files and command-line overrides of those files. This approach and its syntax is modeled on those adopted by the SExtractor photometry package (Bertin \& Arnouts 1996) and will be intuitively familiar to those experienced with SExtractor. This approach allows for fast experimenting with parameter settings and enables efficient scripting.

\section{GENERATION OF MODEL FLUXES AND MAGNITUDES}\label{sec.flux-generation}

The process of generating model SEDs usually begins with unattenuated rest-frame spectra. These input spectra can be either empirical (e.g., Coleman, Wu, \& Weedman 1980) or synthetic (e.g., Bruzual \& Charlot 1993, 2003) and can represent galaxies, AGN, individual Galactic stars, or other objects.  \SEDfit\ does not provide a built-in ability to mix spectra to, e.g., create more complex star formation histories than those already provided in the spectral libraries or to add emission lines to spectra, such as the Bruzual \& Charlot (2003) library, that lack them (but see, e.g., Yabe et al.\ 2009). It is left up to the user to combine or modify spectra if so desired and then save them in the format readable by \SEDfit.  This approach allows maximum flexibility in what models can be fit and \SEDfit's only requirements are that the input spectra follow a standardized format and are given in units of power per frequency interval as a function of wavelength in angstroms, $L_\nu{(\lambda)}$.  This system of units reflects the tendency of high-$z$ optical/IR astronomy to use AB magnitudes (which are defined as logarithms of power per frequency interval\footnote{Recall that $m_{AB} = -2.5 \log [f_\nu / \mathrm{(erg \ s^{-1} cm^{-2}  Hz^{-1}}] - 48.60$  (Oke 1974), or $m_{AB} =  -2.5 \log (f_\nu / \mathrm{nJy}) + 31.4$.}) but to work as a function of wavelength rather than frequency. Then, starting from $L_\nu{(\lambda)}$, several steps are needed to arrive at the spectral energy densities in the observer's frame, $f_\nu(\lambda)$. These steps are performed by the \SEDfit\ program \makesed\ as described below.  

\subsection{Attenuation by interstellar dust}

First, each input spectrum $L_\nu(\lambda)$ is modified to simulate the effect of attenuation due to interstellar dust in the model galaxy.  This attenuation can be expressed as 
\begin{equation}\label{eq.dust-attenuation}
L'_{\nu}{(\lambda)}= L_{\nu}(\lambda) \cdot 10^{-0.4 E(B-V) k(\lambda)}, 
\end{equation}
where $L'$ and $L$ are the attenuated and unattenuated spectral energy distributions, $E(B-V)$ is the color excess whose value controls the overall amount of attenuation, and $k(\lambda)$ is the dust law which defines the profile of attenuation as a function of wavelength.  At present \makesed\ supports the following dust laws: the Calzetti (1997) and Calzetti et al.\ (2000) starburst dust laws, the Fitzpatrick (1986) Milky Way, LMC, and 30 Doradus dust laws, and the Prevot et al.\ (1984) SMC law.  The user specifies one of the dust laws and one or more $E(B-V)$ value(s) to be included in the calculation of the model grid.

\subsection{Attenuation by intergalactic gas}

Next, the spectral energy distribution is attenuated further to account for the stochastic effects of absorption due to intergalactic gas lying between the object being simulated and the observer. At present two intergalactic attenuation prescriptions are included in \makesed:  Madau (1995) and Inoue et al.\ (2005).  The attenuation operation can be expressed as
\begin{equation}\label{eq:intergalactic}
L''_{\nu}{(\lambda)} = L'_{\nu}{(\lambda)} \cdot e^{-\tau_{eff}(\lambda, z)}, 
\end{equation}
where  $L''$ is the SED attenuated by intergalactic gas, $L'$ is the dust-attenuated SED of eq.~\ref{eq.dust-attenuation}, and $\tau_{eff}$ is the effective optical depth as specified by Madau (1995) or Inoue et al.\ (2005).  This optical depth, $\tau_{eff}$, depends on the redshift of the source because the absorbers' column density increases with distance and the absorption wavelengths are subject to the cosmological redshifts of the absorbers.  

The user can chose one of the  two intergalactic attenuation prescriptions and also specify one or more attenuation strengths. The default absorption strength represents a typical line of sight through the Universe, but sightlines that are less or more crowded can also be simulated. Additionally, since the attenuation is redshift-dependent, one or more redshifts --- representing the redshift(s) of the object(s) being simulated --- need to be specified.

\subsection{Cosmological redshifting and dimming}\label{sec.redshift_and_dimming}

The next step is to apply the effects of cosmological distance to the spectrum and thus transform it into the observed frame flux density distribution. This operation can be expressed as 
\begin{equation}
f_{\nu}{(\lambda)} = \frac{L''_{\nu}{(\lambda/(1+z))}}{4\pi D_L^2/(1+z)}, 
\end{equation}
where $D_L$ is the luminosity distance.  The $\lambda/(1+z)$ expression in the numerator indicates that the spectrum's wavelength scale has been redshifted from the original, rest-frame wavelength to one at the desired redshift. The user can chose a desired cosmology by specifying the values for $\Omega_M$, $\Omega_\Lambda$, and $H_0$, while the same single or multiple redshift values are used as for the intergalactic attenuation of eq.~\ref{eq:intergalactic}.

\subsection{Generating model magnitudes}

As the result of the above steps, \makesed\ contais a grid of redshifted and attenuated spectral density distributions.  These SEDs can be saved to a file as a grid of spectra, or can be further processed to generate broadband SEDs by averaging (sometimes called "convolving" in the field) the spectra through system transmission curves, which can be provided by the user and can/should include the filter transmission profile, detector efficiency, and other wavelength-dependent throughput characteristics of the telescope+instrument system. For photon-counting systems, such as CCDs and near-IR arrays, the procedure can be expressed as 
\begin{equation}\label{eq.fmi}
f_{m,i} = \frac{\int f_{\nu}(\lambda) \lambda T_i(\lambda) d\lambda}{ \int \lambda T_i(\lambda) d\lambda}, 
\end{equation}
where $T_i$ is the system transmission curve for the $i^{th}$ filter and $f_{m,i}$ is the model's observer-frame flux density through that filter (the subscript $m$ refers to ``model" as distinct from the observed ``data" flux densities to be encountered in \S\ref{sec.fitting}). The $\lambda$ factor in both numerator and denominator accounts for the fact that the number of photons per unit flux varies with wavelength.  The user can specify one or more transmission curves to be used; a large number of common transmission curves is already supplied with \makesed\ and more can be generated by the user as simple ASCII files. If the input spectrum $L_\nu$ is in units of erg/s/Hz then the output of \makesed\ is in AB magnitudes, 
\begin{equation}
m_{AB, i} = -2.5\log f_{m,i} - 48.60, 
\end{equation}
(Oke 1974).

\subsection{Outputs}

The main end product of \makesed\ is a file that contains one galaxy model per line, with each model consisting of a number of $m_i$ magnitudes along with associated parameters such as redshift, $E(B-V)$ value, galaxy stellar mass, etc.  The number of models in the grid is $N_{models} = N_{E(B-V)} \times N_{z} \times N_{spectra}$, where $N_{E(B-V)}$ is the number of color excess steps sampled, $N_z$ is the number of redshift steps, and $N_{spectra}$ is the number of input spectra (the $F_\nu$) that were used. These input spectra carry with them information such as star formation history, metallicity, etc., and the user can provide as few or as many input spectra as are desired. A typical \makesed\ output file may include 221 spectra (the standard number provided for a given metallicity and star formation history by the GISSEL package of Bruzual \& Charlot 2003), several hundred redshift steps, and several tens of $E(B-V)$ steps. 

This grid of models produced by \makesed\ can be used for exploring the expected magnitudes and colors of high-redshift objects, but more pertinently here it serves as one of the two ingredients (the other being the observational data) in determining the redshifts or physical properties of observed high-$z$ galaxies, as described in \S\ref{sec.fitting}. 

In addition to the model magnitudes grid, \makesed\ can produce a file that contains the full redshifted and attenuated spectra before the integration through system transmission curves (i.e., the end product of \S\ref{sec.redshift_and_dimming}).  Such spectra are not used directly in SED fitting (\S\ref{sec.fitting}) but are useful for, e.g., visualizing the SEDs of distant galaxies, as is illustrated in the left panels of Figs.~\ref{fig.fitexample}  and \ref{fig.ULfitexample} (\S\ref{sec.examples}).

\section{SED FITTING}\label{sec.fitting}

\SEDfit\ uses a maximum-likelihood approach to determine the best-fitting model and to establish confidence regions in parameter space.  These tasks are performed by the program \fitsed. As its inputs \fitsed\ takes the magnitudes model grid produced by \makesed\ (\S\ref{sec.flux-generation}) and a catalog file containing the observed magnitudes or flux densities. The format of the catalog file is flexible as long as it contains one object per line. Operationally, \fitsed\ determines the best-fitting model for a given object by comparing the object's flux densities with the model flux densities which are obtained from the model grid file.  The model that is the most likely, in the sense that it gives the lowest \chisq\ value, is deemed to be the best-fitting model.

\subsection{Determining the goodness-of-fit for a model}

When fitting an object, \fitsed\ properly accounts for both detected and observed but undetected fluxes. When all the photometric data are detected flux measurements, the \fitsed\ compares a model with the observed data and determines the quality of the fit by means of the usual \chisq\ statistic, 
\begin{equation}\label{eq.chisq.body}
\chi^2 = \sum_i \left( \frac{f_{d,i} - s  f_{m,i}}{\sigma_i} \right)^2, 
\end{equation}
where $f_{m,i}$ is the model flux density through the $i^{th}$ bandpass, $f_{d,i}$ is the observed flux density that same bandpass, and $\sigma_i$ is the uncertainty in the observed flux density.  Note that this uncertainty could include both the claimed observational photometric error and an additional term that captures an estimate of systematic uncertainty.  \SEDfit\ does in fact allow the user to specify a $\sigma_{i, sys}$ for each bandpass, in which case $\sigma_i = (\sigma_{i, phot} + \sigma_{i, sys})^{1/2}$.

The quantity $s$ in Eq.~\ref{eq.chisq.body} is the flux scaling between the model and the data which is determined analytically (Sawicki 2002) as 
\begin{equation}\label{eq.s.body}
s = \sum_{i} \frac{f_{d,i} f_{m,i}}{ \sigma_i^2} \Bigg/ \sum_{i} \frac{f^2_{m,i} }{ \sigma_i^2}.
\end{equation}
The derivations of eq.~\ref{eq.chisq.body} and ~\ref{eq.s.body} are reviewed in the Appendix. 

Equation~\ref{eq.chisq.body} yields the correct result when the object is detected in all the observed bandpasses but becomes ambiguous when one or more of the observed bandpasses yield such a non-detection.  However, even non-detections can be useful or even critical (identification of Lyman break galaxies provides an obvious example in this respect) in constraining models or ruling out portions of parameter space.  Consequently, rather than ignoring the information offered by such non-detections, or adopting the ad-hoc approach of assigning arbitrary fluxes to the non-detections, the current implementation of \SEDfit\ properly accounts for such data by using a modified version of the \chisq\ method. The derivation of this modified \chisq\ statistic is given in the Appendix and yields
\begin{eqnarray}\label{eq.chisqUL.body}
\chi^2 & = & \sum_i \left( \frac{f_{d,i} - s  f_{m,i}}{\sigma_i} \right)^2 \nonumber \\
& - & 2\sum_j \ln \int_{- \infty}^{f_{lim, j}} \exp \left[ -\frac{1}{2} \left( \frac{f - s  f_{m,j}}{\sigma_j} \right) ^2 \right]  df.
\end{eqnarray}
The first sum on the right-hand side in eq.~\ref{eq.chisqUL.body} refers to detected bandpasses, indexed with $i$, while the second sum accounts for those bandpasses, indexed with $j$, that yielded upper limits; here $f_{lim, j}$ is the 1-$\sigma$ flux detection treshold in the $j^{th}$ such bandpass, which \fitsed\ calculates from the $n$-$\sigma$ values (where $n$=3, 5, etc.) given in the data catalog. Note that when all bandpasses yield detections, the second term in eq.~\ref{eq.chisqUL.body} drops out and eq.~\ref{eq.chisqUL.body} reduces to the familiar eq.~\ref{eq.chisq.body}, as expected.  Whereas the flux scaling $s$ could be determined analytically for the case when all bandpasses yielded detections (eq.~\ref{eq.s.body}), no such exact analytic expression exists for the case when non-detections are present.  For this reason, the integral in eq.~\ref{eq.chisqUL.body} needs to be evaluated numerically, making the fitting of objects with non-detections significantly more computationally expensive than that of objects that are detected in all bandpasses. Nevertheless, non-detections can be critical in ruling out parts of parameter space (see \S~\ref{sec:ex.ULfitting} for an example) and thus the additional computational cost is often worthwhile\footnote{Were all flux measurements reported, even those below a nominal detection threshold, it would be possible to use eq.~\ref{eq.s.body} in all situations. However, it is usually the case in astronomy that only limits are reported when flux measurements are below an assigned threshold, and the approach taken by eq.~\ref{eq.chisqUL.body} provides a way to include that information in the fit.}.

\subsection{Identifying the best-fit model}\label{sec.finding-best-fit}

The program \fitsed\ searches the model grid to find the model that minimizes \chisq\ (as defined by eq.~\ref{eq.chisqUL.body}).  Two search algorithms are implemented in the standard release of \fitsed\ and the user is free to choose one or the other. 

The simpler and most robust,``brute-force" fitting approach computes \chisq\ for every model in the grid and then unambiguously identifies the one with the lowest \chisq\ value. This approach is robust but --- because of the need to calculate \chisq\ for every model in the grid --- not particularly efficient. However, in many applications it is sufficiently fast while its robustness against effects such as secondary minima as well as its ability to produce complete \chisq\ maps of the model space are its advantages. 

The second available option is to perform a downhill search of the model parameter space starting from a random position in the model grid. Because the downhill search can become lodged in local minima in parameter space, \fitsed\ can repeat the downhill search several times per object, starting with a new random position each time. The user specifies the number of searches per object; experiments show that $\sim$10 repeats per object are typically sufficient to ensure that the absolute minimum is found. Because of its speed, this ``downhill search with random repeats" option is preferable when extremely large sets of objects are to be fitted, or --- because of the computationally-expensive nature of evaluating non-detections (eq.~\ref{eq.chisqUL.body}) --- when many objects with upper limits are present. 

Additional search algorithms (such as simulated annealing) have been implemented within \fitsed\ but it was found that the two standard approaches (the brute-force calculation for every model in the grid and the downhill search with random repeats) are adequate in the vast majority of situations; consequently at present only these two algorithms are included in the standard distribution of \SEDfit. 

The model with the lowest \chisq\ value is by definition the most likely to have produced the observed data and is adopted as the ``best-fitting'' model. The values of the parameters associated with this model (such as redshift, age, spectral type, etc.) are then adopted as the best-fitting parameter values.

\begin{figure*}
\begin{center}
\includegraphics[width=0.95\textwidth]{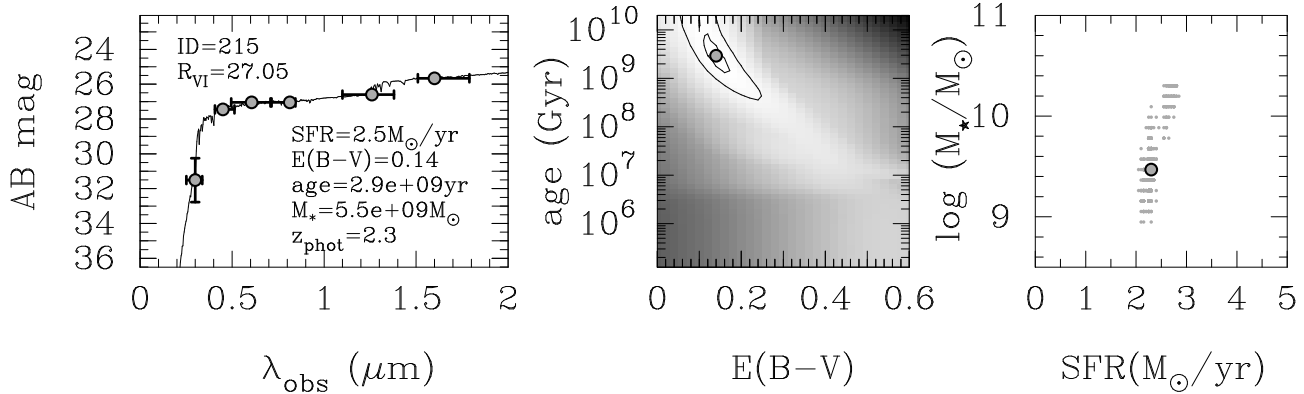} 
\caption{
\label{fig.fitexample} 
An example of SED fitting of a Lyman Break Galaxy in the Hubble Deep Field.  This faint galaxy ($R$=27.05) has no known spectroscopic redshift (redshift is a free parameter of the fit) and in this example has been fit with a suite of solar metallicity, constant star formation rate models from the Bruzual \& Charlot (2003) library. The left panel shows the photometric data points and the best-fit model spectrum (shown as a solid line), with the best-fit model's parameters listed in the figure. The middle and right panels show the location of the best-fit model in parameter space along with an indication of the associated uncertainties. The middle panel shows confidence regions determined from \chisq\ maps: the darker shading indicates higher \chisq\ values and the contours correspond to 1 and 2 $\sigma$ confidence.  The right panel shows the best-fit model along with 1000 Monte Carlo re-fits of the data; the distribution of these 1000 realizations gives an indication of the probability distribution of the model parameters; although this is not done here, formal confidence regions could be generated from these Monte Carlo results by converting them into a two-dimensional histogram and plotting density thresholds.
}
\end{center}
\end{figure*}

\subsection{Determining the uncertainties}

Uncertainties on the values of parameters in the best-fitting model can be established by \fitsed\ in one of several ways.   

The first way is to make use of the Monte Carlo simulations feature of \fitsed.  Here, \fitsed\ refits each object a large, user-specified number of times (e.g., 200 or 1000), but with each realization the photometry in each band is randomly perturbed from the actual value by drawing from a Gaussian distribution with standard deviation dictated by the object's photometric uncertainty in that band.  Note that for operational convenience the perturbation is performed on the model photometry; this is mathematically equivalent to perturbing the observed data and is consistent with the fundamental principle that underlies the $\chi^2$ formalism. The distribution in parameter space  of the best fits to the perturbed photometry realizations then defines the uncertainty ranges on the parameters of the true best-fit model.  \SEDfit\ reports the uncertainties on each parameter in its tabulated output, and the individual Monte Carlo results can also be saved to a file for establishing confidence contours (``error contours") in multi-parameter space.  The Monte Carlo-based uncertainties work when all the photometric bands are detected as well as in the case when some bands are undetected. 

An alternative, faster way, that can be used when the full grid of models has been fit (i.e., the ``brute-force" fit, \S\ref{sec.finding-best-fit}) is to use the \chisq\ map produced during the fitting process.  The offset between each model's \chisq\ value from the \chisq\ of the best-fitting model, $\Delta$\chisq, is a measure of probability of the corresponding model. On the basis of this offset the one-dimensional uncertainties on each parameter as well as full confidence contours in multi-parameter space can be established. The $\Delta$\chisq\ level that corresponds to a desired confidence level for the number of free parameters available can be taken from theory (e.g., Press et al.\ 2007) or can be established empirically for a given dataset using a representative sample of Monte Carlo simulations. 

 \subsection{Outputs}\label{sec.outputs}

The program \fitsed\ produces one standard and several optional output files.  The standard output file lists, for each object in the input catalog, the best-fitting model and its associated parameters and their one-dimensional uncertainties. The user can specify which best-fit model  parameters to report in the standard output file and also which parameters from the input data catalog to propagate into that file. The three optional outputs are:  a file that contains the fit results for all the Monte Carlo realizations; a \chisq\ map for each object in the input file; and a file that contains the shell-style command-line input which, when supplied to \makesed, will result in the production of the redshifted and dust-attenuated spectra that correspond to the best-fit models.   All these optional outputs can be used to illustrate the quality of the SED fit; some examples of the uses of such outputs are given in the figures in \S\ref{sec.examples}.

\section{EXAMPLES OF APPLICATIONS}\label{sec.examples}

This section illustrates the use of \SEDfit\ by means of three examples: (1) the basic SED fitting of a high-$z$ galaxy, (2) SED fitting that makes use of photometric non-detections, (3) two-dimensional SED mapping of high-$z$ galaxies, and (4) simulations of photometric redshift performance.  Of course, \SEDfit\ is not restricted to just these  four specific tasks and many additional examples of \SEDfit\ applications may be found in the papers listed in \S\ref{sec.introduction}. 

\subsection {Simple SED fitting of a distant galaxy}\label{sec:ex.single}

The determination of quantities such as stellar mass, extinction, age, etc.\ constitutes \SEDfit's main raison d'\^etre.  As an illustration, an example of the results of the fitting of a high-$z$ galaxy is given here. Figure~\ref{fig.fitexample} shows the photometric properties of a faint color-selected Lyman Break Galaxy in the Hubble Deep Field and illustrates various aspects of \SEDfit's output, including the best-fitting spectral model (left panel) and confidence regions in parameter space (middle and right panels) determined in two different ways.  

Although, for simplicity, Fig.~\ref{fig.fitexample} presents the results for a single galaxy, \SEDfit\ is designed to routinely process not just individual objects but catalogs of galaxies. Further examples of such work can be found in Sawicki \& Yee (1998), Yabe et al.\ (2009), and Sawicki (2012).

\subsection{SED fitting including non-detections}\label{sec:ex.ULfitting}

\begin{figure*}
\begin{center}
\includegraphics[width=0.95\textwidth]{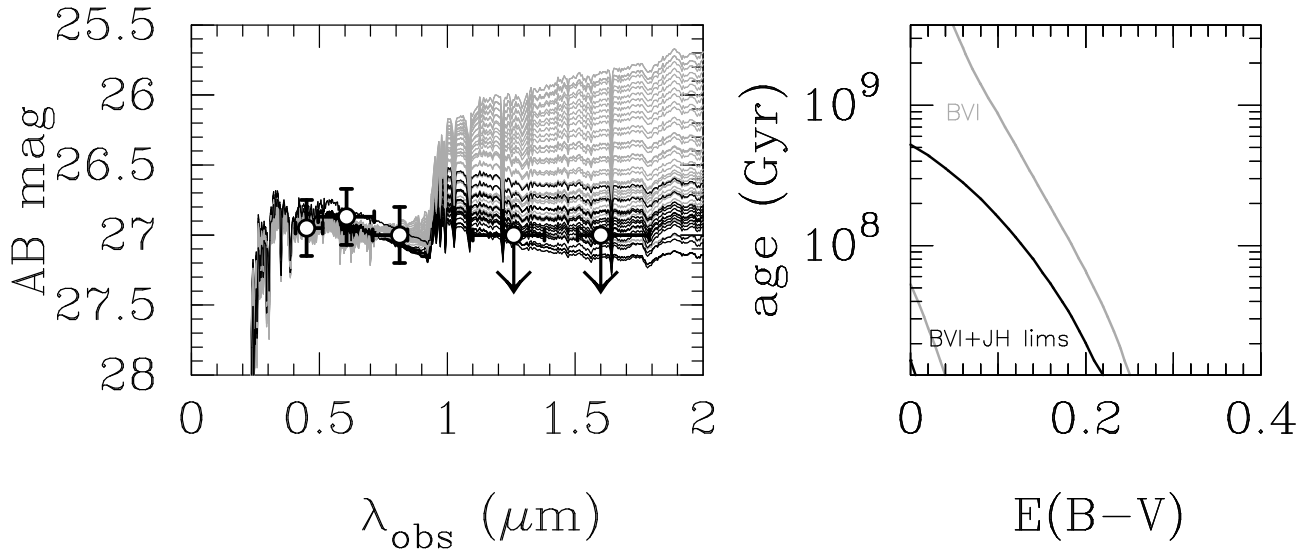} 
\caption{
\label{fig.ULfitexample} 
 An example of SED fitting that uses non-detections to constrain the fit.  The object is a galaxy with redshift fixed at $z=1.5$, photometric detections in the $BVI$ bands and upper limits in $J$ and $H$ (indicated in the left panel with downward-pointing arrows at the 1$\sigma$ flux density limits) fit using Bruzual \& Charlot (2003) constant SFR models.  A sampling of models that give acceptable fits to the detected data (i.e., $BVI$ only) is shown with light-colored curves in the left panel.  Models that give acceptable fits once non-detections are included in the fit are shown with black curves.  The right panel shows regions of age-\ebv\ parameter space that are permitted by the fits:  the light contour encloses models allowed when only detections are used in the fit, while the black contour shows the constraints when non-detections are included. The contours show 68\% confidence regions in both cases. The inclusion of non-detections clearly improves the constraints on the parameter space, here enabling the placement of meaningful constraints on the age of the stellar population. }
\end{center}
\end{figure*}

\SEDfit's ability to include non-detections in the fitting process is illustrated in Fig.~\ref{fig.ULfitexample}. In this example the issue at stake is constraining the age of a stellar population of a \z=1.5 object.  The left panel of Fig.~\ref{fig.ULfitexample} shows both observations that yielded detections ($BVI$) and those that resulted only in upper limits ($JH$). 

When only the detected points are used in the fit, the data altogether fail to constrain the age of the object:  the full range of ages between very young ($<10^7$ yr) and that of the age of the Universe at the object's redshift is allowed, as is shown by the gray contour in the right panel of Fig.~\ref{fig.ULfitexample}. In contrast, when the non-detections are included in the fit (using the full eq.~\ref{eq.chisqUL.body}), a more useful constraint on the age is possible: at the 68\% confidence level, the age is younger than $5\times10^8$ yr (black contour).

The formal maximum-likelihood fit (eq.~\ref{eq.chisqUL.body}) gives a statistically correct way to incorporate the information contained in the non-detections in a way that a ``$\chi$-by-eye" approach of inspecting the models in the left panel of Fig.~\ref{fig.ULfitexample} does not allow.  The fact that some permissible models (black curves) in the left panel of the figure appear higher than the 1$\sigma$ $JH$ limits is simply because such models are indeed permitted by the data.  The most tempting ``$\chi$-by-eye" analysis here would have missed these older but permissible models to give very low but unrealistic constraints on the age. In contrast, the proper maximum-likelihood treatment that uses Eq.~ \ref{eq.chisqUL.body} gives constraints that are both more accurate, automatic, and reproducible.

\subsection{Spatially resolved SED fitting}\label{sec:ex.2d}

\SEDfit\ can also be used to perform spatially-resolved, pixel-by-pixel SED fitting when the target galaxies are sufficiently resolved in the images.  Operationally, the procedure is a simple modification of the procedure used to fit a catalog of individual objects (\S\ref{sec:ex.single}). In this approach, each pixel of a galaxy is fit as if it were an individual object and then the resulting lists of best-fit values are reassembled into parameter maps (e.g., Abraham et al.\ 1999).  

\begin{figure}
\begin{center}
\includegraphics[width=8cm]{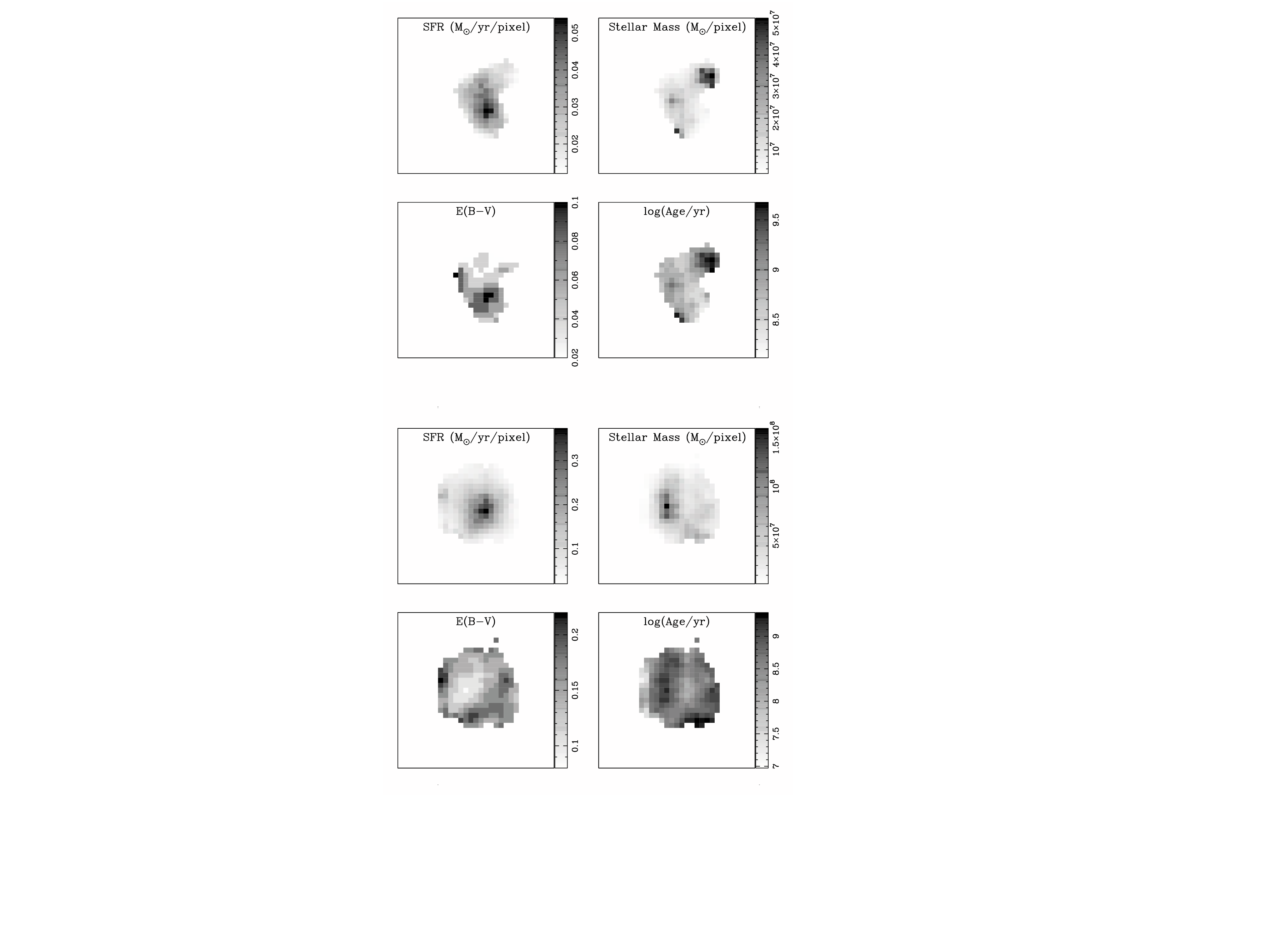}
\caption{
\label{fig.spatially-resolved} 
Best-fit parameter maps for two Hubble Deep Field $z$$\sim$2.3 Lyman Break Galaxies whose individual pixels were fit using \SEDfit.  Each panel is 30 pixels, or 1.2\arcsec, on the side, which corresponds to $\sim$10~kpc at these redshifts. These maps suggest that in both galaxies unobscured older stars are offset from younger, more dust-obscured regions of recent star formation (Palmer 2010).  }
\end{center}
\end{figure}

Figure~\ref{fig.spatially-resolved} illustrates the results of this procedure applied to two Lyman Break Galaxies in the Hubble Deep Field.  Here, HST \hdffilters\ images of spectroscopically-confirmed \zs 2.3 galaxies have been smoothed to a common PSF to guard against artificial color gradients, their individual pixels fit using \SEDfit, and the results reassembled into spatial maps of SFR, stellar mass, color excess \ebv, and age (Palmer 2010).  The resulting maps reveal interesting spatial offsets between regions of ongoing star formation and more quiescent regions of older stars.

\subsection{Simulation of photometric redshifts}\label{sec:ex.photz_sim}

\begin{figure}
\includegraphics[width=8cm]{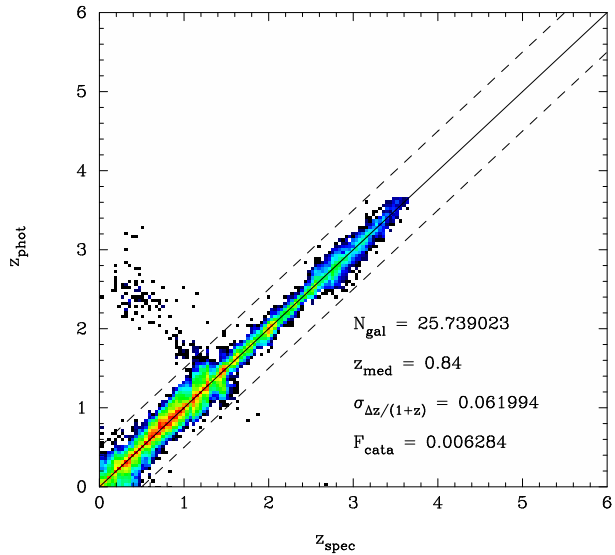} 
\caption{
\label{fig.zzsims} 
Photometric redshift simulations for the Euclid dark energy mission performed using \SEDfit.  The intensity of color in the figure corresponds to the number of galaxies in each $z_{phot}$--$z_{spec}$ bin, with brighter colors corresponding to more objects. The numbers in the lower right of the figure indicate (top to bottom) number of galaxies per square arcminute, the median redshift, the photometric redshift bias, and the fraction of catastrophic redshift failures, defined here as the fraction of objects outside the dashed lines (see \S\ref{sec:ex.photz_sim} and Sorba \& Sawicki 2011 for additional details).  Analysis of simulations like these provides the ability to fine-tune the design of photometric redshift surveys. }
\end{figure}

\SEDfit\ can be used to estimate photometric redshifts to distant galaxies (e.g., Sawicki et al.\ 1997; Sawicki \& Mall\'en-Ornelas 2003; Sorba \& Sawicki 2010) and also to perform simulations of photometric redshift performance (e.g.\ Sorba \& Sawicki 2011).  Figure~\ref{fig.zzsims} shows an example of such simulations: the specific case here shows the expected fidelity of the photometric redshift performance of the proposed Euclid dark energy mission when augmented by the addition of $U$-band photometry to the planned Euclid filter set.  In this example, photometry for several hundred thousand simulated galaxies was generated using \makesed, drawn from a range of spectral templates and using a realistic redshift distribution. These galaxies' photometry was then perturbed according to the expected photometric uncertainties. Finally, the ``observed'' galaxies were fit using \fitsed\ with the same spectral templates as used in their generation.  Figure~\ref{fig.zzsims} shows the ``observed'' photometric redshift vs.\ the input ``spectroscopic'' redshifts; because the same spectral template set was used to generate the artificial galaxies as to SED-fit them, the scatter in the ``observed'' redshifts from the diagonal line results purely from photometric uncertainties in the data. Analysis of this type can be useful when designing photometric redshift surveys.

\section{DISCUSSION}\label{sec.discussion}

Given observational photometric data and a set of model spectra, SED fitting can produce useful insight into the properties of distant objects.  In this regard, the most commonly-sought parameters are photometric redshifts, galaxy stellar masses, amount of interstellar extinction, dust-corrected star formation rates, and ages of stellar populations.  The technique has now been widely used and forms a standard part of the toolbox of extragalactic astronomy.  

Despite the apparent simplicity of SED fitting and the clear usefulness of the physical properties that it purports to derive from pure photometric data, it is important to use the technique with caution since there are a number of issues that can bias the results.  Some of these limitations are discussed in the following paragraphs. 

First, no matter how good the spectral models, the observational data need to be of the right quality.  This means not only a sufficiently large and well-chosen set of photometric bandpasses, but also accurate photometry, since otherwise systematic errors in, e.g., photometric calibration, will propagate into systematic errors in the estimated parameters such as stellar masses and star formation rates. In this respect it is crucial to realize that accurate colors (flux density ratios) are far more important than the absolute calibration of the overall spectrum:  for example, a 10\% offset in overall, across-all-bands calibration will result in only a 10\% offset in stellar mass and star formation rate, but a 10\% offset in \emph{relative} photometry between bands can result in far worse and even catastrophic inaccuracies in the fitted quantities.  Careful matching of photometric apertures across bandpasses to ensure accurate flux density ratios is thus essential, including procedures to account for PSF differences between bands. 

A second set of limitations stems from the inability of SED fitting to constrain certain key properties.  A well-known limitation (e.g., Sawicki \& Yee 1998; Papovich et al.\ 2001) is the inability to constrain star formation histories from the typical broadband photometry available for distant objects.  The degeneracy in star formation histories propagates into degeneracies in other fit parameters, most notably the star formation rates. There is no good way around this issue given limited photometric data and thus one simply has to be aware of this limitation when interpreting fitting results.  In this respect it is perhaps wise to perform relative --- rather than absolute --- measurements, comparing results across similarly-selected samples of galaxies fit with a single set of models rather than trusting the absolute results of the best-fit parameters.  

A further limitation is the SED fitting technique's focus on the most luminous stellar population present in the object being fitted.  This limitation most commonly manifests itself in the inability to properly account for an older stellar population when a young population is present.  The limitation stems from the fact that old stellar populations are faint in rest-frame UV compared to young populations that are dominated by luminous high-mass stars.  Old stellar populations, if they exist, will thus often be missed in the glare of even a sprinkling of bright young stars. This effect can be critical since such old, undetected underlying stellar populations can contain significant masses of stars, potentially several times the mass of the young detected populations. 

An additional limitation stems from the input spectra used to generate the model templates: the results of the SED fitting will be only as good as these spectra.  While a number of excellent stellar population models is now available, significant differences between these models remain and suggest that our understanding of stellar populations is still not perfect. The resulting estimates of galaxy physical parameters will thus carry associated systematic errors. In light of such differences it is perhaps wise to again rely on relative comparisons between fit galaxy properties (masses, star formation rates, etc.) rather than on the absolute values of these properties. 

A related point is that SED fitting will produce a best-fitting model regardless of whether that model is or is not a good match to the observations. It is therefore wise to inspect the goodness-of-fit parameter, \chisq, to ensure that the best-fitting model is indeed a well-fitting model (the two are not synonymous!).  Further, irrespective of how good a \chisq\ is produced, even a well-fitting best-fit model may not be a true reflection of reality since degeneracies in fitting space exist and the SED-fitting procedure may preferentially choose a slightly better-fitting, but ultimately incorrect model over a worse-fitting but correct one.  Consideration of such degeneracies should be an important aspect of interpreting SED-fitting results.

Despite systemic limitations such as those discussed above, SED fitting has proven to be a useful and popular technique in extragalactic research. It provides us with a way to estimate key quantities of interest, such as photometric redshifts, galaxy stellar masses, amount of extinction due to dust, dust-corrected star formation rates, and ages of stellar populations; and it lets us do so for a large number of objects from a single set of common observations. 
The technique has served us in this role for over a decade now and will likely continue to be a key tool in our toolbox well into the coming  era of ever more powerful observational facilities and ever larger datasets.

\vspace{5mm}

I am grateful to Huan Lin who contributed code to an earlier version of \SEDfit, and Ikuru Iwata, who implement the Inoue et al.\ (2005) intergalactic opacity function in the present version of the software; to Michael Palmer, Bobby Sorba, Kiyo Yabe, and Suraphong Yuma for requests that resulted in extending the capabilities of this software; to Bobby Sorba for allowing me to show the results of our Euclid photo-z simulations in Fig.~\ref{fig.zzsims} and to Michael Palmer for letting me display results from his undergraduate honours thesis in Fig.~\ref{fig.spatially-resolved}.   I also thank 
Liz Arcila-Osejo, 
Anneya Golob, 
Michael Palmer, 
Taro Sato, 
Barbara Sawicki, 
Jerzy Sawicki, 
Bobby Sorba, 
Kiyo Yabe, and 
Suraphong Yuma
and the anonymous referee
for comments that improved the quality of this manuscript. Finally, this software has evolved over a number of years, and thus funding from several agencies has contributed to its development: I gratefully acknowledge support from the Natural Sciences and Engineering Research Council (NSERC) of Canada, the Canadian Space Agency (CSA), NASA, an International Long-Term Visitorship from the Japan Society for the Promotion of Science (JSPS), and the Atlantic Computational Excellence Network (ACEnet).

\appendix

\section{THE MAXIMUM LIKELIHOOD FORMALISM FOR SED FITTING WITH UPPER LIMITS}\label{sec.fitting-formalism}

\SEDfit's model-fitting program, \fitsed\ uses a maximum-likelihood approach to identify the best-fitting model and determine confidence regions in parameter space.  This appendix first reviews the derivation of the usual \chisq\ formula as applied to the particular case of SED fitting and then uses a similar approach to derive a variant of \chisq\ that is appropriate when some of the data are non-detections. 

The standard way to approach the development of a maximum-likelihood fitting method is to assume that a given model underlies the observed data and the data are drawn from that model but are perturbed by Gaussian errors (see, e.g., Press et al.\ 2007).  In other words, the question one asks is ``what is the probability that the observed data could have occurred if the data were drawn from the model that's being considered?"

\subsection{The case when the object is detected in all bands}

\begin{figure*}
\begin{center}
\includegraphics[width=8cm]{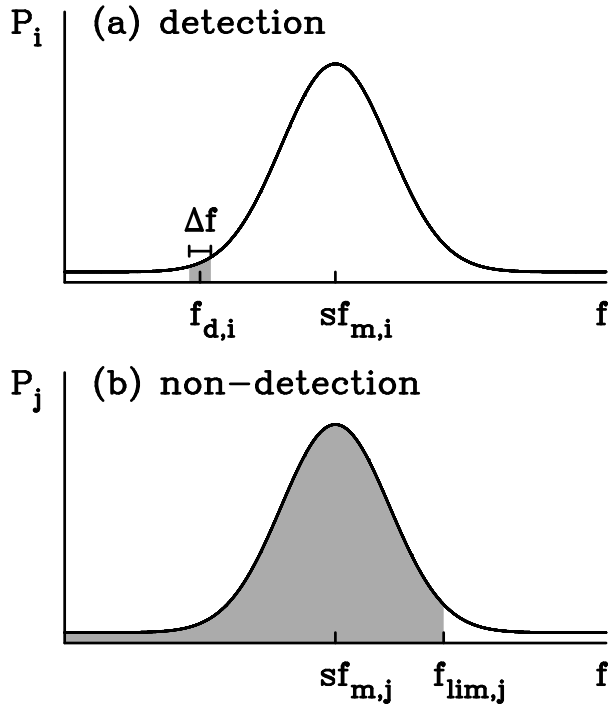} 
\caption{
\label{fig.probabilities} 
Probability that an observation is drawn from a given model.  The top panel is for the case of a detection, which leads to the well-known standard \chisq\ goodness-of-fit estimator.  The bottom panel is the equivalent diagram for the case of a non-detection.  In both cases the model predicts the flux density value $sf_{m}$ and the Gaussian curves show the probability distribution under the influence of Gaussian perturbation of the flux. The shaded regions (in practice infinitessimally thin in the case of a detection) correspond to the probability that the random Gaussian uncertainties have perturbed the true flux to result in the observed detection at $f_{d,i}$ (top panel) or non-detection below the detection threshold $f_{lim,j}$ (bottom panel).}
\end{center}
\end{figure*}

If the object has been detected in all photometric bands then the approach outlined above produces the well-known \chisq\ formalism, as is shown below (see, e.g., \S 15.1 of Press et al.\ 2007, for a general \chisq\ derivation).  For a single band $i$, the probability that a set of observations has been produced from a given model is 
\begin{equation}\label{eq.probability_i}
P_i \propto \exp \left[ -\frac{1}{2} \left( \frac{f_{d,i} - s  f_{m,i}}{\sigma_i} \right) ^2 \right] \Delta f , 
\end{equation}
where $f_{d,i}$ is the observed (data) flux density in the $i$th band, $\sigma_i$ is its standard deviation (i.e., uncertainty), and $s  f_{m,i}$ is the model flux density in the same band.  The quantities involved are illustrated in Fig.~\ref{fig.probabilities}(a).   Note that we have separated the model flux density into two components:  the quantity $s$ is the normalization of the model, which, for a given model, is the same across all the bands $i$, while $f_{m,i}$ contains the information about the spectral shape of the model.  This separation allows us to considerably speed up the numerical handling of the calculation, as is shown below (eq.~\ref{eq.s}).  In practice, having $f_{m,i}$ distinct from the normalization $s$ is also convenient because it is $f_{m,i}$ that is produced from the spectral models (eq.~\ref{eq.fmi}), while $s$ contains the information about the normalization adjustment that best brings such a model into match with the observed data and gives the multiplicative scaling that links the model's scaleable parameters --- such as luminosity, stellar mass, and star formation rate --- with those in the galaxy under consideration. 

The overall probability of the data in \emph{all} the observed bandpasses being drawn from the model in question is then proportional to the product of the individual probabilities, 
\begin{equation}\label{eq.product}
P \propto \prod_i P_i.
\end{equation}
Next, taking the logarithm of eq.~\ref{eq.product} gives
\begin{equation}\label{eq.probabilities_i}
P \propto \ln \prod_i P_i = -\frac{1}{2} \sum_i \left( \frac{f_{d,i} - s  f_{m,i}}{\sigma_i} \right)^2  +  \sum_i \ln \Delta f .
\end{equation}
The second sum in eq.~\ref{eq.probabilities_i} is a constant and thus maximizing the probability is just equivalent to minimizing the first sum, \begin{equation}\label{eq.chisq}
\chi^2 = \sum_i \left( \frac{f_{d,i} - s  f_{m,i}}{\sigma_i} \right)^2.
\end{equation}
The derivation thus far has followed the standard derivation of the \chisq\ formula as a maximum likelihood estimator with the assumption of normally distributed errors (e.g., Press et al.\ 2007). 

The normalization factor $s$ of the best-fitting model can be derived analytically and doing so can considerably speed up the numerical calculation of \chisq. This analytic solution can be achieved by minimizing the \chisq\ of eq.~\ref{eq.chisq} by taking the partial derivative $\partial \chi^2 / \partial s$ and setting it to zero (Sawicki 2002).  Doing so yields the expression
\begin{equation}\label{eq.s}
s = \sum_{i} \frac{f_{d,i} f_{m,i}}{ \sigma_i^2} \Big/ \sum_{i} \frac{f^2_{m,i} }{ \sigma_i^2},
\end{equation}
which exactly gives the appropriate scaling $s$ of the model. Identifying the model with the smallest \chisq\ from among the model suite gives the most likely (i.e., best-fitting) model in the case when the object is detected in all the observed bands. 

\subsection{The case of upper limits}\label{sec.upperlimits}

For the case where the object is not detected in one or more of the observed bands we can proceed as above but with replacing detections with upper limits in eq.~\ref{eq.probability_i}--\ref{eq.chisq}.  In the case of a non-detection in the band $j$, the probability that the observation in that band is drawn from a given model is 
\begin{equation}
P_j \propto \int_{-\infty}^{f_{lim, j}} \exp \left[ -\frac{1}{2} \left( \frac{f - s  f_{m,j}}{\sigma_j} \right) ^2 \right]  df,
\end{equation}
where $f_{lim,j}$ is the upper limit (i.e., sensitivity) of the observation in the $j$th band. In analogy to eq.~\ref{eq.product}, the probability of the data (detections and non-detections) across all the observed bandpasses being drawn from the given model is then the product of the individual probabilities, 
\begin{equation}
P \propto \prod_i P_i \prod_j P_j.
\end{equation}
The subscripts $i$ indicate those bands which have yielded detections, while the subscripts $j$ are for the bands with non-detections.  The equivalent of eq.~\ref{eq.probabilities_i} then is
\begin{equation}
P \propto  -\frac{1}{2} \sum_i \left( \frac{f_{d,i} - s  f_{m,i}}{\sigma_i} \right)^2  +  \sum_i \ln \Delta f 
   + \sum_j \ln \int_{-\infty}^{f_{lim, j}} \exp \left[ -\frac{1}{2} \left( \frac{f - s  f_{m,j}}{\sigma_j} \right) ^2 \right]  df, 
\end{equation}
and, in analogy with eq.~\ref{eq.chisq}, maximizing the likelihood that the observed dataset (detections and non-detections) is drawn from the given model is equivalent to minimizing the quantity 
\begin{equation}\label{eq.chisqUL}
\chi^2  = \sum_i \left( \frac{f_{d,i} - s  f_{m,i}}{\sigma_i} \right)^2
-  2\sum_j \ln \int_{-\infty}^{f_{lim, j}} \exp \left[ -\frac{1}{2} \left( \frac{f - s  f_{m,j}}{\sigma_j} \right) ^2 \right]  df.
\end{equation}
Note that if all the observed bands yielded detections then the last term drops out and eq.~\ref{eq.chisqUL} reduces to the form of eq.~\ref{eq.chisq}, as one would expect. 

For computational convenience, the integral in eq.~\ref{eq.chisqUL} can be recast in terms of the error function, ${\rm erf}(x)= (2/\sqrt{\pi}) \int_0^x e^{-t^2} dt$, so that
\begin{equation}\label{eq.chisqULerf}
\chi^2  = \sum_i \left( \frac{f_{d,i} - s  f_{m,i}}{\sigma_i} \right)^2
  -2\sum_j \ln \left\{
  \sqrt{\frac{\pi}{2}}\sigma_j     
  \left[ 1 + {\rm erf} \left(
  \frac{f_{lim, j} - s  f_{m,j}}{\sqrt{2}\sigma_j}
\right) \right]
  \right\}.
\end{equation}
Finding the most likely model requires evaluating eq.~\ref{eq.chisqULerf} to identify $\chi^2_{min}$, the smallest value of $\chi^2$. In the case when all the bands yield detections finding $\chi^2_{min}$ can be accelerated by optimizing the model scaling factor $s$ using eq.~\ref{eq.s}. In the present case of non-detections there is no simple equivalent of eq.~\ref{eq.s}.  One approach is then to find the $s$ that's optimal for a given model by numerically exploring eq.~\ref{eq.chisqULerf}.  Alternatively, in analogy with eq.~\ref{eq.s}, setting $\partial \chi^2 / \partial s = 0$ gives the condition for the optimal $s$ for a given model:
\begin{equation}\label{eq.sUL}
\sum_i  \left( \frac{f_{d,i} - s  f_{m,i}}{\sigma_i} \right)
	\left( \frac{f_{m,i}}{\sigma_i} \right)
- \sqrt{\frac{2}{\pi}}\sum_j
	\frac
		{
			{f_{m,j} \exp 
			\left\{
				-
					\left[
						(f_{lim,j} - s f_{m,j}) / 
						\sqrt{2}\sigma_j
					\right]^2 	
			\right\} 
			}	
		}
		{
			{\sigma_j 
			\left\{
				1 + {\rm erf}
					\left[							
							(f_{lim,j} - s f_{m,j}) / \sqrt{2} \sigma_j
						\right]
			\right\}}
		}
=0.
\end{equation}
Finding the root of eq.~\ref{eq.sUL} then gives the optimal $s$ for the model. This root can be obtained numerically using a variety of root-finding methods. The present implementation of \SEDfit\ uses simply numerically searches for a model's optimal $s$ in eq.~\ref{eq.chisqULerf}.




\clearpage






%

\end{document}